\documentclass[preprint,showpacs,preprintnumbers,amsmath,amssymb]{revtex4}

\newcommand\rf[1]{(\ref{eq:#1})}
\newcommand\lab[1]{\label{eq:#1}}

\newcommand\br{\begin{eqnarray}}
\newcommand\er{\end{eqnarray}}
\newcommand\be{\begin{equation}}
\newcommand\ee{\end{equation}}

\newcommand\bc{\begin{center}}
\newcommand\ec{\end{center}}

\newcommand\grad{\nabla}

\newcommand{\bib}[1]{\bibitem{#1}}
\newcommand\PRL[3]{\textsl{Phys. Rev. Lett.} \textbf{#1}, #3 (#2)}
\newcommand\NPB[3]{\textsl{Nucl. Phys.} \textbf{B#1}, #3 (#2)}

\newcommand\PRD[3]{\textsl{Phys. Rev.} \textbf{D#1}, #3 (#2)}

\newcommand\PLB[3]{\textsl{Phys. Lett.} \textbf{#1B}, #3 (#2)}
\newcommand\CQG[3]{\textsl{Class. Quantum Grav.} \textbf{#1}, #3 (#2)}
\newcommand\JMP[3]{\textsl{J. Math. Phys.} \textbf{#1}, #3 (#2)}
\newcommand\PTP[3]{\textsl{Prog. Theor. Phys.} \textbf{#1}, #3 (#2)}

\newcommand\AoP[3]{\textsl{Ann. of Phys.} \textbf{#1}, #3 (#2)}

\newcommand\PR[3]{\textsl{Phys. Reports} \textbf{#1}, #3 (#2)}

\newcommand\IJMPA[3]{\textsl{Int. J. Mod. Phys.} \textbf{A#1}, #3 (#2)}
\newcommand\IJMPD[3]{\textsl{Int. J. Mod. Phys.} \textbf{D#1}, #3 (#2)}

\newcommand\MPLA[3]{\textsl{Mod. Phys. Lett.} \textbf{A#1}, #3 (#2)}

\newcommand\PHSA[3]{\textsl{Physica} \textbf{A#1}, #3 (#2)}

\begin{document}


\title{ Gravitational Theory with a Dynamical Time}

\author{E.I. Guendelman}%
\email{guendel@bgumail.bgu.ac.il}, 
\affiliation{%
Department of Physics, Ben-Gurion University of the Negev \\
P.O.Box 653, IL-84105 ~Beer-Sheva, Israel}%

\begin{abstract}
A gravitational theory involving a vector field $\chi^{\mu}$ , whose zero component has the properties of a dynamical time, is studied.
The variation of the action  with respect to $\chi^{\mu}$ gives the covariant conservation of an energy momentum tensor $ T^{\mu \nu}_{(\chi)}$. 
Studying the theory in a background which has killing vectors and killing tensors we find appropriate shift symmetries of the
field $\chi^{\mu}$ which lead to conservation laws. The energy momentum that is the source of gravity  $ T^{\mu \nu}_{(G)}$
is different but related to  $ T^{\mu \nu}_{(\chi)}$ and the covariant conservation of  $ T^{\mu \nu}_{(G)}$ determines in general
the vector field $\chi^{\mu}$. When $ T^{\mu \nu}_{(\chi)}$ is chosen to
be proportional to the metric, the theory coincides with the Two Measures Theory, which has been studied before in relation to the Cosmological Constant Problem. When the matter model consists of point particles, or strings, the form of $ T^{\mu \nu}_{(G)}$, solutions for $\chi^{\mu}$ are found.
For the case of a string gas cosmology, we find that the Milne Universe can be a solution, where the gas of strings does not curve the spacetime
since although $ T^{\mu \nu}_{(\chi)} \neq 0$, $ T^{\mu \nu}_{(G)}= 0$, as a model for the early universe, this solution is also free of the horizon problem. There may be also an application to the "time problem" of quantum cosmology.

\end{abstract}


\maketitle

\newpage

\section{Introduction and Summary}

In the standard approach to dynamical systems, the conservation of energy, in the particle mechanics case, or the local conservation
of an energy momentum tensor in the field theory case, are derived results rather than starting points. For example, the conservation of energy can be derived from the time translation invariance principle.

In this paper we will see however that conservation of energy (in a particle mechanics example) and local conservation of an energy momentum tensor (in the field theory case) can be starting points rather than derived results. This as we will see is interesting because a theory constructed along these lines contains new a dynamical variable,  which in the field theory case is a vector field $\chi^{\mu}$ whose zero component can be interpreted as representing a dynamical time.

Indeed in the formulation of this kind of field theory, energy densities and momentum densities are canonically conjugated variables to  $\chi^{0}$ and $\chi^{i}$ respectively, which is what we expect from a dynamical time (here represented by 
$\chi^{0}$). Also supporting the interpretation of  $\chi^{0}$ as representing a dynamical  time is given by the interesting fact that in the presence of killing vectors and killing tensors, symmetries involving
shifts of  $\chi^{\mu}$, which lead to conservation laws can be found, so that all the components of the vector $\chi^{\mu}$ and the zero component in particular have the properties of a coordinate, although it is a field.

In the fully covariant formulation and when the metric is also dynamic rather than a background, one can define the gravitional energy momentum tensor,$ T^{\mu \nu}_{(G)}$ , which is the object which (up to a proportionality constant) equals the Einstein tensor. $ T^{\mu \nu}_{(G)}$ is different but related to  $ T^{\mu \nu}_{(\chi)}$ and the covariant conservation of  $ T^{\mu \nu}_{(G)}$ determines in general
the vector field $\chi^{\mu}$. The theory deviates from GR for strong gravitational fields and when $ T^{\mu \nu}_{(\chi)}$ is chosen to
be proportional to the metric, it coincides with the Two Measures Theory, which has been studied before in relation to the Cosmological Constant Problem \cite{TMT}, \cite{TMT1}.

We will discuss these kind of models for some specific matter models, for example for a system of point particles and for a system of strings. 

When the matter model consists of point particles, or strings, the form of $ T^{\mu \nu}_{(G)}$ and  solutions for $\chi^{\mu}$ are found.
For the case of a string gas cosmology, we find that the Milne Universe can be a solution, where the gas of strings expands but it does not curve the spacetime since although $ T^{\mu \nu}_{(\chi)} \neq 0$, $ T^{\mu \nu}_{(G)}= 0$. There is a big bang of the matter, without an associated initial curvature singularity since the space is indeed flat. As a model for the early universe, this solution is also free of the horizon problem.

Finally we will discuss the use of the vector field $\chi^{\mu}$ in the context of quantum gravity and how this could resolve the "problem of time" in quantum cosmology.

\section{A Simple Mechanical Example}
In order to see how these ideas are implemented, we start with the simplest possible example, the dynamics of a particle in a potential $V(x)$
where the energy conservation is implemented from the begining.

We start by defining the energy functional that we want to be conerved as 
\be
 \frac{1}{2}m v^2 + V(x)
\lab{energy functional}
\ee
where $v=\frac{dx}{dt}$.

We then consider the action principle

\be
S = \int dt L = \int dt\frac{da}{dt}(\frac{1}{2}m v^2 + V(x)) 
\lab{action}
\ee
here the role of the dynamical variable $a$ is to implement the conservation of energy as an equation of motion from the start,
indeed the variation of this action with respect to $a$ gives

\be
\frac{d}{dt}(\frac{1}{2}m v^2 + V(x)) = 0
\lab{eq.from a}
\ee
which is integrated to give

\be
\frac{1}{2}m v^2 + V(x) = E = constant
\lab{energy conservation}
\ee
of course we can differentiate \rf{energy conservation} with respect to time which gives

\be
m \frac{dv}{dt}v + \frac{dV(x)}{dx}v = 0
\lab{eq. for v}
\ee

One can check complete consistency indeed with other equations, from the variation of the action with respect to $x$ 
we get
\be
m\frac{d}{dt}( \frac{da}{dt}\frac{dx}{dt}) = \frac{dV(x)}{dx} \frac{da}{dt}
\lab{yet another eq.}
\ee

It is interesting to notice that the action \rf{action} is time independent and there is therefore a conservation law
arising from this, which is

\be
H = \frac{da}{dt} p_a + \frac{dx}{dt}p_x - L  = \frac{dx}{dt}p_x = m v^2 \frac{da}{dt} = constant
\lab{Hamiltonian}
\ee
so that  $\frac{da}{dt} = c/v^2 $, which inserted into \rf{yet another eq.} gives

\be
m\frac{d}{dt}( \frac{c}{v}) = \frac{dV(x)}{dx} \frac{c}{v^2}
\lab{final eq.}
\ee
which indeed gives $m\frac{d^2 x}{dt^2 } = - \frac{dV(x)}{dx} $.

We see the basic pattern arising: the conservation of energy functional is implemented by the introduction of an additional
variable (here $a$), then there is an additional conservation law, that of $H$, which can be used to determine $a$.
We will see how this generalizes to generally coordinate invariant models.

\section{The Generally Coordinate Invariant Case}
Let us consider a $4-D$ case where we will imposse the conservation of a symmetric energy momentum tensor $ T^{\mu \nu}_{(\chi)}$ by introducing the term in the action

\be
S_{(\chi, T)} = \int d^4x\sqrt{-g}\chi_{\mu ;\nu}T^{\mu \nu}_{(\chi)}
\lab{S chi, T equation}
\ee
where

\be
\chi_{\mu ;\nu} = \frac{\partial\chi_{\mu}}{\partial x^{\nu}} - {\Gamma^\lambda_{\mu \nu}}\chi_{\lambda}
\lab{cov.der.}
\ee
if we assume $ T^{\mu \nu}_{(\chi)}$ to be independent of $\chi_{\mu}$ and having
$\Gamma^\lambda_{\mu \nu}$ being defined as the Christoffel Connection Coefficients, then the variation with respect to $\chi_\mu $
gives the local conservation of $ T^{\mu \nu}_{(\chi)}$.

Notice the interesting fact that the energy density is the  canonically conjugated variable to  $\chi^{0}$, which is what we expect from a dynamical time (here represented by 
$\chi^{0}$).

An interesting particular case is obtained when  $ T^{\mu \nu}_{(\chi)}$ is taken to be of the form

\be
T^{\mu \nu}_{(\chi)} = g^{\mu \nu} L_1
\lab{proportional case}
\ee
then introducing in eq. \rf{S chi, T equation} gives

\be
S_{(\chi, T)} = \int d^4x  \partial_\mu (\sqrt{-g} \chi^\mu ) L_1 = \int d^4x  \Phi L_1
\lab{TMT}
\ee

There could be also a "regular" piece, so we obtain a total action of the form,

\be
S =   \int d^4x  \Phi L_1 + \int d^4x \sqrt{-g}L_2
\lab{TMT-total}
\ee
where we can take for $L_1$ an arbitrary function of the fields, curvature, etc. and the same goes for $L_2$. Notice that  \rf{TMT} is like a regular contribution to a standard gravity theory except that instead of $\sqrt{-g} $, in that part of the action
the measure of integration is the total derivative $\Phi= \partial_\mu (\sqrt{-g} \chi^\mu )$. These kind of contributions have been considered in the Two Measures Theories, which are of interest in connection with the Cosmological Constant Problem.
Notice that we can use as dynamical variable in the action principle $\Psi^\mu = \sqrt{-g} \chi^\mu $, in which case it is apparent that 
$\Phi= \partial_\mu  \Psi^\mu $ is independent of the metric. In \cite{TMT}, \cite{TMT1} $\Phi= \partial_\mu  \Psi^\mu $ has been considered where $\Psi^\mu$ is taken as $\Psi^\mu =\epsilon^{\mu \nu \alpha \beta}A_{\nu \alpha \beta} $ and where $A_{\nu \alpha \beta}$ can be a fundamental field or itself a composite of four scalars (this does not change the results). Comelli develops also a similar approach using a vector field to describe the modified measure \cite{TMT2}.

For \rf{proportional case}, the covariant conservation of  $ T^{\mu \nu}_{(\chi)}$ gives rise to the equation $L_1 = M= constant$ (a similar constraint is not achieved in the approach of ref. \cite{TMT2} since there the vector field is assigned a potential in addition to appearing in the measure of integration in the action). 

The Unimodular theory \cite{unimodular}, when expressed in its generally coordinate invariant form \cite{unimodularGCIF} is indeed a special case of the TMT class of theories. Indeed, in \cite{unimodularGCIF} an action of the TMT form \rf{TMT-total} is studied, where  
$L_1= \frac{2\Lambda}{16\pi G}$ and $L_2= -\frac{1}{16\pi G}(R+2\Lambda)$, in the notation of \cite{unimodularGCIF} $\Phi= \partial_\mu T^\mu$, here $\Lambda$ is taken as a dynamical variable, but the equation (as in any TMT type theory) $L_1= M=constant$, forces $\Lambda$ to be a constant, the variation of $\Lambda$, gives a relation between the two measures, $\Phi = \partial_\mu T^\mu = \sqrt{-g}$. We see that of course the vector $T^\mu$ is undetermined, only its four divergence is defined, in particular we could chose apparently the "gauge" $T^0=0$ if we wish, nevertheless on the assumption $T^i \rightarrow 0$ (this could also be taken as a gauge choice, since only the divergence of $T^\mu$ is really determined) at spacial infinity or if the universe is spatially closed (then an integral of a spatial derivative must vanish), we obtain that the 4-D integration ($ \int d^4x $) of the two sides of $ \partial_\mu T^\mu = \sqrt{-g}$   means that $\int d^3x T^0$  has the interpretation of the volume of spacetime (taking as a boundary condition that at $t=0$ the volume is taken as zero). The $T^0$ variable by itself, which is what potentially determines a "dynamical time" in the unimodular theory (the analogous to our $\sqrt{-g} \chi^0 $) is however impossible to determine in this context. 
Also, even in the more general TMT case, only the 4-divergence of the vector is defined, so once again the determination of its zero component is impossible. In the models we will 
discuss here, we will see that a dynamical time can be extracted locally (not just some integrals of it).

The unimodular theory is a equivalent to GR
at the classical level, except that the cosmological constant is an integration constant. The TMT is not equivalent to GR in its generic form, neither is the Gravitational Theory with a Dynamical Time developed in this paper, indeed we see that
the form \rf{S chi, T equation} generalizes the Two Measures Theory approach. However, it is reasonable to think that when looking at the vacuum sector of the theory, the form \rf{proportional case} could be taken, but for more generic cases we have to stay with \rf{S chi, T equation}.

Notice that for \rf{S chi, T equation}, the shift symmetry of  $ T^{\mu \nu}_{(\chi)}$ holds

\be
T^{\mu \nu}_{(\chi)} \rightarrow T^{\mu \nu}_{(\chi)} + \Lambda g^{\mu \nu} 
\lab{shift symmetry}
\ee
where $\Lambda$ is a constant. Since then the integrand in the action changes by the addition of a total derivative.

This means that if the matter is coupled through its energy momentum tensor as in \rf{S chi, T equation}, a process of redefition of the energy
momentum tensor, such as \rf{shift symmetry} will not affect the equations of motion. Of course such type of redetinition of the energy momentum tensor is exactly what is done in the process of normal ordering in Quantum Field Theory for example.

\section{Killing Vector Symmetries}
One interesting feature of 
$S_{(\chi, T)}$ as given by eq. \rf{S chi, T equation} is that if we are considering the situation where the metric
$ g_{\mu \nu} $ is a background ($\chi^\mu $ here will not be a background, since we will require it to transform), that is , is fixed and there are Killing vectors and Killing tensors, then conservation laws that follow from 
shift symmetries involving $\chi^\mu $ follow.

Let us start considering the situation where there is a Killing vector $k_\mu $

\be
k_{\mu ;\nu}+ k_{\nu ;\mu} = 0
\lab{killing vector}
\ee

Then obviously the following is a symmetry of $S_{(\chi, T)}$ as given by eq. \rf{S chi, T equation} ,

\be
\chi_\mu \rightarrow \chi_\mu +k_{\mu}  
\lab{killing vector symmetry}
\ee

Straightforward application of Noether`s theorem leads us to the existence of the conserved current
$ j^\mu$ satisfying $\partial_\mu j^\mu = 0$ and given by
\be
j^\mu = \sqrt{-g}  T^{\mu \nu}_{(\chi)} k_{\nu} 
\lab{killing vector current}
\ee

This is the standard form of the conserved quantity related to the existence of a Killing vector like \rf{killing vector}.
Notice however that the symmetry transformation is expressed as a transformation of the field $\chi_\mu $, no coordinate 
transformation is invoked and indeed the coordinates are kept unchanged. The symmetries described here are different to what is usually considered, which are diffeomorphism transformations that leave invariant the metric \cite{Wald}.

\section{Formulation of Point Particle Models}
 We look at specific matter models and we start with the simplest,
the point particle case.

In this case, we can take for $T^{\mu \nu}_{(\chi)}$ the following expression
\be
  T^{\mu \nu}_{(\chi)} = m\int d\lambda \frac {\delta^4(x-x(\lambda))}{\sqrt{-g}}\frac {dx^\mu}{d\lambda} \frac {dx^\nu}{d\lambda}
\lab{particle e.m. tensor}
\ee
inserting this expression in the form for $S_{(\chi, T)}$ as given by eq. \rf{S chi, T equation} , we get

\be
S_{(\chi, T)} = m\int d\lambda \chi_{\mu ;\nu}\frac {dx^\mu}{d\lambda} \frac {dx^\nu}{d\lambda}\\
= m\int d\lambda \frac {dx^\mu}{d\lambda} (\frac {d\chi_\mu}{d\lambda}-\Gamma^\sigma_{\mu \nu} \frac {dx^\nu}{d\lambda}\chi_\sigma)
\lab{S chi, particle}
\ee
we see then that the variation of \rf{S chi, particle} gives rise to the geodesic equation for $x^\mu$

\be
\frac {d^2x^\mu}{d\lambda^2}+\Gamma^\mu_{\alpha \beta} \frac {dx^\alpha}{d\lambda}\frac {dx^\beta}{d\lambda} =0
\lab{particle geodesic}
\ee
of course this only reproduces the well known result that the conservation of energy momentum (which is the eq. that the
variation of $\chi_{\mu}$ enforces) is equivalent to the conservation of the energy momentum tensor.

Notice also that if we calculate the variation of the action \rf{S chi, particle} with respect to
$x^\mu$ we will also find a geodesic equation , but now with $\chi_{\mu ;\nu}$  playing the role of a metric. The consistency of this with
\rf{particle geodesic} gives a non trivial constraint on $\chi_{\mu}$ as we will discuss latter in some particular examples.
Coupling of particles was considered in the TMT theory, see for example the third reference of \cite{TMT1} in connection to models with dilatons and avoiding the fifth force problem. There the coupling was only through  the divergence$\Phi= \partial_\mu (\sqrt{-g} \chi^\mu )$ however, which differs from the approach here, but we will study in the future the idea of scale invariant coupling to point particles as done in that paper but adopting the more generic framework considered here. 
\section{Killing Tensor Symmetries for Point Particle Models}
It is very interesting that the point particle model described in the previos section allows for shift symmetries
associated with the existence of a Killing tensor. For  example given a symmetric tensor $K_{\mu \nu} $ satisfying the Killing tensor condition

\be
 K_{(\mu \nu; \alpha)}=0
\lab{killing tensor cond.}
\ee
where $()$ means symmetrization of the indices   $\mu, \nu$ and $\alpha $.
If \rf{killing tensor cond.} is satisfied, then it is a straightforward excercise to show that 
the following is a world line symmetry of \rf{S chi, particle}, up to the integral of a total derivative,

\be
 \chi_{\mu} \rightarrow \chi_{\mu} + \epsilon K_{\mu \nu}\frac {dx^\nu}{d\lambda}
\lab{killing tensor symmetry}
\ee
($\epsilon$ being a constant number) 
indeed the variation of the action \rf{S chi, particle} under \rf{killing tensor symmetry}
is,

\be
\delta S_{(\chi, T)} = m\int d\lambda (\epsilon K_{\mu \alpha} \frac {dx^\alpha}{d\lambda} )_{;\nu}\frac {dx^\mu}{d\lambda} \frac {dx^\nu}{d\lambda}= m\int d\lambda \epsilon K_{\mu \alpha} (\frac {dx^\alpha}{d\lambda})_{;\nu} \frac {dx^\mu}{d\lambda} \frac {dx^\nu}{d\lambda}
\lab{killing tensor conservation derivation1}
\ee
where we have used in the second step the Killing tensor condition. The right hand side of \rf{killing tensor conservation derivation1}
is a total derivative and equals to
\be
\delta S_{(\chi, T)} = \frac{m}{2}\int d\lambda  \frac{d}{d\lambda}(\epsilon K_{\mu \alpha} \frac{dx^\alpha}{d\lambda} \frac{dx^\mu}{d\lambda})
 = \frac{m}{2}\int d\lambda  \frac{dx^\nu}{d\lambda}\frac{\partial}{\partial x^\nu}(\epsilon K_{\mu \alpha} \frac{dx^\alpha}{d\lambda} \frac{dx^\mu}{d\lambda})
 \lab{killing tensor conservation derivation2}
\ee

When performing the ordinary derivative with respect to $x^\nu$ in \rf{killing tensor conservation derivation2}, which acts on a scalar,
which means that it is as well a covariant derivative, we can act therefore inside the parenthesis 
in each individual term with covariant derivatives. Using the Killing tensor condition and the symmetry of $ K_{\mu \alpha}$ we obtain therefore the identity between 
\rf{killing tensor conservation derivation2} and \rf{killing tensor conservation derivation1}, which proves \rf{killing tensor symmetry} is a symmetry up to the integral of a total derivative and the Noether charge, which is conserved and that is associated to 
\rf{killing tensor symmetry} is,
 
\be
 Q=  K_{\mu \nu} \frac {dx^\mu}{d\lambda} \frac {dx^\nu}{d\lambda}
\lab{killing tensor conservation law}
\ee
  
This conservation law in the presence of a Killing tensor has been discussed in \cite{killing tensor ref}, but 
the symmetry associated to \rf{killing tensor conservation law} was discussed in a Hamiltonian framework, i.e., not as
a symmetry of an action, but rather of a certain Hamiltonian. Our discussion of symmetries is in the context of
the symmetries of an action and of the straightforward application of Noether`s theorem. Another difference is that the 
Hamiltonian symmetries discussed in \cite{killing tensor ref} involve $x^\mu$, while in our case $x^\mu$ is not affected
and instead the field $\chi_\mu $ is transformed.

The  generalization to higher symmetric Killing tensor symmetries is straightforward, for example, given a three index symmetric Killing tensor 
$K_{\mu \nu \kappa }$
satisfying $K_{(\mu \nu \kappa ; \alpha)}=0$, we find the cooresponding symmetry of \rf{S chi, particle} which is now
$\chi_{\mu} \rightarrow \chi_{\mu} + \epsilon K_{\mu \nu \kappa}\frac {dx^\nu}{d\lambda}\frac {dx^\kappa}{d\lambda}$
and the corresponding Noether charge $Q=  K_{\mu \nu \kappa} \frac {dx^\mu}{d\lambda} \frac {dx^\nu}{d\lambda}\frac{dx^\kappa}{d\lambda}$
and so on for higher rank Killing tensors.

\section{Gravitational Equations and The Gravitational Energy Momentum Tensor}
In the very simple mechanical example of section II, where in addition to the conservation law enforced
by the equation of motion of the variable $a$, there was the additional conserved quantity $H$.
We will see now that in the fully dynamical gravitational theory, this pattern is repeated and in addition to
the energy momentum tensor $T^{\mu \nu}_{(\chi)}$, whose conservation law is implemented due to the equation of
motion due to the field $\chi_\mu $, there will be also the energy momentum tensor which is the source of gravity
and that is also going to be conserved due to the Bianchi identities.

In order to see how this work, we continue looking at the simplest matter model,
the point particle case, but now we will consider gravity as a dynamical field rather than a background. Of course there are
many ways to achieve this, but for the purpose of starting with the simplest possibility, we add a term in the action proportional
to the scalar curvature to \rf{S chi, T equation}, 
, that is, we consider the full action as
\be
S = \int d^4x\sqrt{-g}\chi_{\mu ;\nu}T^{\mu \nu}_{(\chi)} + \frac {1}{16\pi G}\int d^4x\sqrt{-g}R
\lab{full action}
\ee
here we assume the "second order formalism", that is $R$ is defined as the usual curvature scalar defined in terms of the Christoffel
connections which are functions of the metric and its derivatives, that is the dynamical variables are the metric $g_{\mu \nu}$ and the vector field $\chi_\mu $. Using the connection as an independent variable (first order formalism) will not be studied here, but will be the subject of future investigations.

The gravitational equations are then (using units where $8\pi G=1$)

\be
G^{\mu \nu}=R^{\mu \nu}-\frac{1}{2}g^{\mu \nu}R =\frac{1}{\sqrt{-g}} \frac{\partial(\sqrt{-g}L_m)}{\partial g_{\mu \nu}}-
\frac{1}{\sqrt{-g}}\frac{\partial}{\partial x^{\lambda}}(\frac{\partial(\sqrt{-g}L_m)}{\partial g_{\mu \nu,\lambda}})
\lab{grav. eq.}
\ee
here $\sqrt{-g}L_m =\sqrt{-g}\chi_{\mu ;\nu}T^{\mu \nu}_{(\chi)}$.
Notice that \rf{grav. eq.} applied directly to a "vacuum" energy momentum of the form
$T^{\mu \nu}_{(\chi)} = \Lambda g^{\mu \nu} $ (where $\Lambda $ is a constant) and therefore
$\sqrt{-g}L_m =\sqrt{-g}\chi_{\mu ;\nu}T^{\mu \nu}_{(\chi)}=\Lambda \sqrt{-g}\chi_{\mu ;\nu}g^{\mu \nu}$ , gives identically zero, in agreement with the basic fact that the theory is not sensitive to the presence of a vacuum piece in 
$T^{\mu \nu}_{(\chi)}$ which is a consequence of the shift the symmetry \rf{shift symmetry} .

Now we will calculate the right hand side of \rf{grav. eq.} in the case where
$\sqrt{-g}\ T^{\mu \nu}_{(\chi)}$ is independent of the metric, having in mind the point particle case
where $T^{\mu \nu}_{(\chi)}$ is given by \rf{particle e.m. tensor} so that the metric dependence in $\sqrt{-g}L_m $ is only through the connection that appears in $\chi_{\mu ;\nu}$.For the calculation of \rf{grav. eq.} the following results are of use,
\be
\frac{\partial\Gamma^\tau_{\lambda \sigma}}{\partial g_{\mu \nu}}= -\frac{1}{2}(g^{\mu \tau}\Gamma^\nu_{\lambda\sigma}+g^{\nu \tau}\Gamma^\mu_{\lambda\sigma})\\
\lab{derivatives1}
\ee
\be
\frac{\partial\Gamma^\tau_{\lambda \alpha}}{\partial g_{\mu \nu,\sigma}}= \frac{1}{4}(g^{\tau \mu}(\delta^{\nu}_{\alpha} \delta^\sigma_\lambda+\delta^{\nu}_{\lambda} \delta^\sigma_\alpha) + g^{\tau \nu}(\delta^{\mu}_{\alpha} \delta^\sigma_\lambda+\delta^{\mu}_{\lambda} \delta^\sigma_\alpha) - g^{\tau \sigma}(\delta^{\mu}_{\alpha} \delta^\nu_\lambda+\delta^{\mu}_{\lambda} \delta^\nu_\alpha))
\lab{derivatives2}
\ee
from these results and using also the covariant conservation of $ T^{\mu \nu}_{(\chi)}$,  we obtain for the \\"gravitational energy momentum tensor", that is  for $G^{\mu \nu}=R^{\mu \nu}-\frac{1}{2}g^{\mu \nu}R =T^{\mu \nu}_{(G)}$ the expression
\be
T^{\mu \nu}_{(G)}= -\frac{1}{2}(\chi^\alpha \grad_{\alpha})T^{\mu \nu}_{(\chi)} - \frac{1}{2} T^{\mu \nu}_{(\chi)}(\grad_{\alpha}\chi^\alpha)
+\frac{1}{2}T^{\nu \sigma}_{(\chi)}\chi^\mu_{;\sigma} +\frac{1}{2}T^{\mu \sigma}_{(\chi)}\chi^\nu_{;\sigma}
\lab{T Grav}
\ee
here $;$ and $\grad$ both represent covariant differentiation. As a simple and fast way to obtain \rf{T Grav} one may first calculate in a locally inertial frame where the
Christoffel symbol and all derivatives of the metric vanish and then go back to an arbitrary frme by converting ordinary derivatives by covariant ones.

We see that the energy momentum tensor $T^{\mu \nu}_{(\chi)}$ determines $T^{\mu \nu}_{(G)}$ but it is not equal to it.
The expression has been derived under the assumption that $\sqrt{-g}\ T^{\mu \nu}_{(\chi)}$ is independent of the metric.
In the point particle model, we could add to $T^{\mu \nu}_{(G)}$ an additional piece proportional to $T^{\nu \sigma}_{(\chi)}$ 
by considering an additional piece in the action of the form
\be
\omega\int d^4x\sqrt{-g} \int d\lambda \frac {\delta^4(x-x(\lambda))}{\sqrt{-g}}g_{\mu \nu}\frac {dx^\mu}{d\lambda} \frac {dx^\nu}{d\lambda}
\lab{additional}
\ee

This is the generalization of the procedure developed in the special case of the TMT, where we allow coupling to the conventional measure
$\sqrt{-g}$ as well a to $\Phi$.

\section{When the vector field can be considered a background?. Flat space and some cosmological cases.}
After looking at the structure of the equations, we see that there are two fundamental objects governing the motion of matter, the metric
$g_{\mu \nu}$ and the vector field $\chi_\mu$.

In the sections related to the shift symmetries that transform $\chi^\mu $, when there are killing vectors and killing tensors, the metric was considered a background but not $\chi^\mu $. Now we will explore whether $\chi^\mu $ could also be considered a background alongside with the metric.

The covariant conservation of  $T^{\mu \nu}_{(G)}$ provides indeed an equation of motion for the vector field $\chi_\mu$, so the vector field $\chi_\mu$ depends rather strongly on the matter content, so in general it cannot be considered as a "background".

There may be exceptions however where this may be possible, but in order to make things well defined, we will assume that in any point in space there is at least some matter, i.e., everywhere $T^{\mu \nu}_{(\chi)} \neq 0 $, for this we could think for example of a continous distribution of particles (fluid) rather than a discrete collection of point like particles. This requirement is made since $\chi_\mu$ is determined from the covariant conservation of $T^{\mu \nu}_{(G)}$ and only if $T^{\mu \nu}_{(\chi)} \neq 0 $  this will give a well defined result for $\chi_\mu$.
Of course going beyond point particles and into field theory, there is usually no such thing as empty space anywhere, since fields are defined at any point in space, even more extreme is the quantum field theory case where empty space is not possible, even in principle due to the quantum fluctuations.

Assuming then that both $\chi_\mu$ and $g_{\mu \nu}$ are both backgrounds means that they are both not modified very much by the precise state of the matter. This imposses a non trivial condition since as we have seen, the equation of motion obtained from the variation of $\chi_\mu$ is the geodesic equation with respect to the metric
$g_{\mu \nu}$, while the variation with respect to $x^{\mu}$ means the particles must move also as geodesics with respect to the "effective metric"
being the symmetric part of $\chi_{\mu ;\nu}$, even if we generalize the model to include a contribution like \rf{additional} the particles must move also as geodesics with respect to the "effective metric" a linear combination of the symmetric part of $\chi_{\mu ;\nu}$ and $g_{\mu \nu}$
as well as being a geodesic with respect to $g_{\mu \nu}$.

If both $g_{\mu \nu}$ and $\chi_\mu$ are going to be backgrounds, not substantially affected by the particles that move through them and arbitrary geodesics are geodesics of both $g_{\mu \nu}$ and of the symmetric part of $\chi_{\mu;\nu}$ , then apparently this can only happen if these two objects are proportional
\be
 \chi_{\mu ;\nu}+ \chi_{\nu ;\mu}= cg_{\mu \nu}
\lab{relation}
\ee
where $c$ is a constant.

Condition \rf{relation} means $\chi_\mu$ is a conformal killing vector of the metric, but not only that since in the right hand side of 
\rf{relation} the metric appears multiplied by a constant factor, which makes it a very special conformal killing vector called "homothetic vector field".

Many cosmological spaces satisfy \rf{relation} where $c$ is a constant, for example, any Robertson Walker space with a power law factor and $k=0$ \cite{flatness}.  In particular
flat space satisfies  \rf{relation} with  $\chi_\mu$  being given by
\be
\chi^{\mu} =\frac{c}{2} x^{\mu}
\lab{homothetic killing vector}
\ee
where $ x^{\mu}$ are the Minkowski flat coordinates.

In this case $\chi^{\nu}_{;\mu}= \frac{c}{2}\delta^{\nu}_{\mu}$ and $T^{\mu \nu}_{(G)}$ as given by \rf{T Grav}
becomes
\be
T^{\mu \nu}_{(G)}=  -\frac{c}{4}(x^\alpha \partial_{\alpha})T^{\mu \nu}_{(\chi)} - \frac{c}{2} T^{\mu \nu}_{(\chi)}
\lab{T Grav-flat}
\ee
the two pieces in the right hand side of \rf{T Grav-flat} are separatelly conserved, since $\partial_{\mu}T^{\mu \nu}_{(\chi)} =0$
and as a consequence of this $\partial_{\mu}((x^\alpha \partial_{\alpha})T^{\mu \nu}_{(\chi)}) =0$. It is interesting to notice that for any integer $n$, $\partial_{\mu}((x^\alpha \partial_{\alpha})^n T^{\mu \nu}_{(\chi)}) =0$ if $\partial_{\mu}T^{\mu \nu}_{(\chi)} =0$ is assumed.

The term $-\frac{c}{4}(x^\alpha \partial_{\alpha})T^{\mu \nu}_{(\chi)}$ looks exotic, but if we exam its contribution to the gravitational mass
of a localized static configuration, we see it gives the rather conventional form,

\be
 -\int d^3 x \frac{c}{4}(x^\alpha \partial_{\alpha})T^{00}_{(\chi)} =-\int d^3 x \frac{c}{4}(x^i \partial_i)T^{00}_{(\chi)}
=\frac{3c}{4}\int d^3 x T^{00}_{(\chi)}
\lab{T Grav-flat-mass}
\ee
where in the last step we have integrated by parts, given that the configuration is localized.  
The total mass is then
\be
 \frac{c}{4}\int d^3 x T^{00}_{(\chi)}
\lab{Tot-mass}
\ee
$c$ redefines the gravitational constant, or alternativelly one can think of this as modifying the mass of the object. Of course, if there is an additional term of the type described by \rf{additional},
this will also add to the total contribution. The "exotic term" can have effects that differentiate the theory from GR in the cosmological setting however, see for this section X.

One can treat in a similar fashion to this case of flat spacetime any space for which a vector field satisfying \rf{relation}, 
for example a Robertson Walker space with a power law factor and $k=0$ ,
\be
 ds^2= -dt^2 + t^{2\alpha}(dx^2+dy^2+dz^2) 
 \lab{Rob.walker}
\ee
then  $\chi_\mu$ can be taken as the vector that satisfies \rf{relation} and then both the metric and $\chi_\mu$ can be taken as backgrounds, 
in the case of \rf{Rob.walker} will be given by
\be
 \chi^\mu= \frac{c}{2}(t, (1-\alpha)x, (1-\alpha)y, (1-\alpha)z) 
 \lab{Rob.walker hom. killing}
\ee
this should be inserted then in our expression for $T^{00}_{(G)}$ \rf{T Grav}.

\section{Application to Strings and other Generalizations}
We can follow the same scheme and consider strings. We start with the form \rf{S chi, T equation}, where $T^{\mu \nu}_{(\chi)}$
will be given by ($a, b$ represent the world sheet coordinates $\sigma ,  \tau$)
\be
T^{\mu \nu}_{(\chi)}=\int \int \partial_a X^{\mu }\partial_b X^{\nu}\gamma^{ab} \sqrt{-\gamma} \frac{\delta^4(x-X(\sigma, \tau))}{\sqrt{-g}}d\sigma d\tau
\lab{string e.m.}
\ee
here $\gamma_{ab}$ is, as in Polyakov approach, an auxiliary world sheet metric which is self consistently determined by its equation of motion and $\gamma = det(\gamma_{ab})$. When we insert \rf{string e.m.} into \rf{S chi, T equation}, and solve for $\gamma_{ab}$, we will find that
$\gamma_{ab}$ is, up to a conformal factor the induced metric on the string, but where the bulk metric (which we project onto the world sheet) is taken as the symmetric part of
$\chi_{\mu ;\nu}$. The equation of motion of $\chi_{\mu}$ on the other hand enforces the covariant conservation of \rf{string e.m.}
where in this conservation law the covariant derivatives are defined with respect to the metric $g_{\mu \nu}$.

The condition under which \rf{T Grav} was derived, i.e., that $\sqrt{-g}\ T^{\mu \nu}_{(\chi)}$is independent of the metric $g_{\mu \nu}$, since $\gamma_{ab}$ is a priori independent of $g_{\mu \nu}$, although it depends on it after solving the equations of motion. Then
when using \rf{string e.m.}, still holds, therefore, we still obtain \rf{T Grav} in this case. Even if we solve for $\gamma_{ab}$, after this replace back into the action and then calculate the variations with respect to the metric, this does not afect the validity of \rf{T Grav}, because
although there is an apparent additional dependence on $g_{\mu \nu}$ through $\gamma_{ab}$, the fact that the variation of $\gamma_{ab}$ extremizes the action guarantees that the dependence of $g_{\mu \nu}$ through $\gamma_{ab}$ does not contribute when calculating the variation of the action with respect to $g_{\mu \nu}$  and therefore anyway we do it,  \rf{T Grav} is correct. 

The formulation of the string case is not only generally coordinate invariant, but also world sheet reparametrization invariant. We could also
study the point particle case in a world line reparametrization invariant way. This can be achieved by working with the following action

\be
S=S_{(\chi, particle)} + \frac {1}{16\pi G}\int d^4x\sqrt{-g}R
\lab{reparametrization invariant}
\ee
where
\be
S_{(\chi, particle)} = \int d^4x\sqrt{-g}\chi_{\mu ;\nu}m\int d\lambda \frac {\delta^4(x-x(\lambda))}{\sqrt{-g}}\frac {dx^\mu}{d\lambda} \frac {dx^\nu}{d\lambda}e + \int d^4x\sqrt{-g} m\int d\lambda \frac {\delta^4(x-x(\lambda))}{\sqrt{-g}}e^{-1}
\lab{reparametrization invariant part.}
\ee
here we have introduced the additional world line auxiliary field $e$ that ensures that world line reparametrization invariance 
$\lambda \rightarrow \lambda^{\prime} = \lambda^{\prime}(\lambda)$, since then $e$ can transform as $e \rightarrow e^{\prime} = \frac {d\lambda^{\prime}}{d\lambda} e$. 

Notice that the second term in the right hand side of \rf{reparametrization invariant part.} does not contain any $g_{\mu \nu}$ dependence
and the first term in the right hand side of \rf{reparametrization invariant part.} obeys all the conditions required for its variation with respect $g_{\mu \nu}$ to to be given by \rf{T Grav}, where  $T^{\mu \nu}_{(\chi)}$ is now given by
\be
  T^{\mu \nu}_{(\chi)} = m\int d\lambda  \frac {\delta^4(x-x(\lambda))}{\sqrt{-g}}\frac {dx^\mu}{d\lambda} \frac {dx^\nu}{d\lambda} e
\lab{particle e.m. rep. inv}
\ee
the additional factor of $e$ makes sure that $T^{\mu \nu}_{(\chi)}$ is now world line reparametrization invariant. The variation of $\chi_{\mu}$
makes sure that $T^{\mu \nu}_{(\chi)}$ is covariantly conserved. The situation with $e$ is similar to the role played by $\gamma_{ab}$ in the string case and once again
the condition under which \rf{T Grav} was derived, i.e., that $\sqrt{-g}\ T^{\mu \nu}_{(\chi)}$is independent of the metric $g_{\mu \nu}$, since $e$ is a priori independent of $g_{\mu \nu}$, although it depends on it after solving the equations of motion. Then
when using \rf{string e.m.}, still holds, therefore, we still obtain \rf{T Grav} in this case. Even if we solve for $e$, after this replace back into the action and then calculate the variations with respect to the metric, this does not afect the validity of \rf{T Grav}, because
although there is an apparent additional dependence on $g_{\mu \nu}$ through $e$, the fact that the variation of $e$ extremizes the action guarantees that the dependence of $g_{\mu \nu}$ through $e$ does not contribute when calculating the variation of the action with respect to $g_{\mu \nu}$  and therefore, once again, anyway we do it,  \rf{T Grav} is correct. 

\section{Self Consistent solutions, The Non Gravitating String Gas Cosmology}
Let us consider \rf{homothetic killing vector}, which is the appropriate solution for $\chi_{\mu}$ given that the space is Minkowski flat space. However, Minkowski flat space will not be a solution if $T^{\mu \nu}_{(G)} \neq 0$. Interestinly enough, according to \rf{T Grav-flat}, we could have non trivial matter content, i.e. $T^{\mu \nu}_{(\chi)}\neq 0$, which does not produce a gravitational effect, i.e., giving
$T^{\mu \nu}_{(G)} = 0$. \rf{T Grav-flat} tells us that $T^{\mu \nu}_{(G)} = 0$ is satisfied provided $T^{\mu \nu}_{(\chi)}$ is a homogeneous function of the Minkowski coordinates $x^{\mu}$ of degree $-2$. 

Let us then consider a form which is a homogeneous function of the Minkowskii coordinates of degree $-2$ and which furthermore has the Lorentz covariant form, 
\be
  T^{\mu \nu}_{(\chi)} = \frac{A}{x_\alpha x^\alpha }  \eta^{\mu \nu} + \frac{B }{(x_\alpha x^\alpha)^2 }x^\mu x^\nu  
\lab{Lorentz cov. form}
\ee
where $x_\alpha x^\alpha =x^\alpha x^\beta \eta_{\alpha \beta}$, $A$ and $B$ are constants. The local conservation law $\partial_{\mu} T^{\mu \nu}_{(\chi)}= 0$ implies then that $B=2A$. Writting \rf{Lorentz cov. form} in a hydrodynamic fashion, introducing the four velocity 
$u^{\mu}=\frac{x^\mu}{(-x_\alpha x^\alpha)^{1/2}}$  and  defining density and pressure as 
\be
  T^{\mu \nu}_{(\chi)} =  u^{\mu}u^{\nu} \rho + p(\eta^{\mu \nu} + u^{\mu}u^{\nu})
\lab{density and pressure}
\ee
we obtain that $\rho = -\frac{3A}{x_\alpha x^\alpha } $ and $p=\frac{A}{x_\alpha x^\alpha }$, that is the equation of state 
$p=-\frac{1}{3}\rho $
is satisfied, which is the case for a string gas for example. Since according to our analysis eq. \rf{T Grav} holds also for strings,
then the consideration of string matter in the context of that equation and therefore also of  \rf{T Grav-flat} is self consistent.

The solution can be cast in a cosmological fashion although the metric is flat and since the energy density and pressure depend only on 
$x_\alpha x^\alpha =x^\alpha x^\beta \eta_{\alpha \beta} =-(x^0)^2 + (x^1)^2+(x^2)^2+(x^3)^2$, it is natural to define a time variable t as 
$-t^2 \equiv -(x^0)^2 + (x^1)^2+(x^2)^2+(x^3)^2$, this can be done by going to the Milne coordinates $t, \psi, \theta,\phi$ related to the Minkowski coordinates by,
\be
 x^0=tcosh\psi,  x^1=tsinh\psi sin\theta cos\phi,  x^2=tsinh\psi sin\theta sin\phi, x^3=tsinh\psi cos\theta 
\lab{Milne coordinates}
\ee
then the metric becomes
\be
 ds^2=-dt^2+ t^2( d\psi^2 + sinh^2\psi(d\theta^2 +sin^2\theta d\phi^2))
\lab{Milne metric}
\ee
which represents a Robertson Walker cosmology with negative spatial curvature and although the metric it is flat spacetime, it represents a non trivial solution of the gravitational equations, with non zero matter content, $\rho = \frac{3A}{t^2 } $ and 
$p=-\frac{A}{t^2} =-\frac{1}{3}\rho$, $t$ defined as 
$-t^2 \equiv -(x^0)^2 + (x^1)^2+(x^2)^2+(x^3)^2$  turns out to be cosmic time. So, homogeneous and isotropic expanding matter obeying the "string gas" or "K-matter" \cite{Kolb} equation of state do not curve the spacetime. If this were to represent a model of the early universe, we will have the big bang at $t=0$,
where all matter will have an infinite density, but no curvature singularities appearing, since the space is Minkowski flat.
This solution is also free of the horizon problem, because cosmologies with a linear dependence in time for the Robertson Walker factor, as we have here, provide an infinite horizon distance (as in the Kolb coasting cosmology \cite{Kolb}).

\section{Discussion, conclusions and prospects}
We have seen that introducing a vector field that enforces the conservation of an energy momentum tensor gives rise to an interesting new gravitational theory. In the case the energy momentum tensor that we choose be conserved is of the form $T^{\mu \nu}_{(\chi)} = g^{\mu \nu}L $,
where $L$ is an arbitrary function of the fields, the conservation of $T^{\mu \nu}_{(\chi)} $ leads to the constraint $L= M =constant$ and the theory reduces to the Two Measures Theory that has been extensively studied in its many variations\cite{TMT}, \cite{TMT1} (a similar constraint is not achieved in the approach of ref. \cite{TMT2} since there the vector field is assigned a potential in addition to appear in the measure of integration in the action).

Going beyond this particular case of the Two Measures Theory, we see that the theory generically gives rise to two non trivial energy momentum tensors which are related but different, $T^{\mu \nu}_{(\chi)} $ the original energy momentum that the vector field enforces its covariant conservation and $T^{\mu \nu}_{G}$, the gravitational energy momentum tensor, so this generalization of the Two Measures Theory could be called the Two Energy Momentum Tensors Theory. 

We have studied this for the case of the point particle and strings as matter models, but of course more matter models could be studied.
In particular it could be interesting to study these ideas in the context of supersymmetry, since the energy momentum tensor has known super partners and one could think that a similar procedure to the studied in this paper could lead also to an additional energy momentum tensors with its related superpartners.

For the additional vector field, its zero component appears to be like of a "dynamical time", this is supported by several facts, first its canonically conjugated momenta is the energy density, second Killing vector and tensor symmetries are simply implemented in terms of shifts of the vector field, so the vector $\chi^\mu $ behaves like the coordinates of spacetime and its zero component like a temporal codinates and finally the explicit solutions show that the zero component of this vector field in a Minkowski background can be taken consistently as the Minkowski time and a similar type of solution for some other cosmological backgrounds.

We expect the "vacuum sector" of the theory to be like that of the TMT, since for this vacuum sector the energy momentum tensors should be proportional to the metric, so the generic theory of this type should preserve  the favorable properties of TMT concerning the cosmological constant problem.

One should also explore the difference between the theories that result when assuming the first order formalism and the second order formalism,
for example allowing the connection coefficients that appear in the definition of $R$ to be independent variables as has been done in the case of the TMT. So far in this paper we have only worked with the second order formalism. Also, one should study the possible incorporation of the idea of scale invariance, as has been done in the context of TMT \cite{TMT1}.

An important subject to be discussed are the observational constraints on the model. As we have seen in section VII, in addition to the "exotic"
new terms that couple the point particles to the vector field $\chi_\mu$, we can always add also the conventional term of the type \rf{additional}, so if necessary the "exotic" terms can be made small as compared to the ordinary ones.
Notice however that the deviations are not seen as deviations of the geodesic eq. since the role of $\chi_\mu$ is precisely to enforce the geodesic equation. The coupling to $\chi_\mu$ changes however the definition of the "gravitational energy momentum tensor", so it has the potential to change the gravitational effects of a point particle, i.e., the gravitational field produced by a point particle. However, even as a source of gravity one will have to work hard to find discrepacies, this is because for example for spherical symmetry, outside a source we must still have a Schwarzschild geometry.

When the matter model consists of point particles, or strings, the form of $ T^{\mu \nu}_{(G)}$  and the self consistent solutions for $\chi^{\mu}$ are found. Furthermore we have seen that these kind of expressions in the cosmological setting can lead to solutions where, although having non trivial matter content, i.e. $T^{\mu \nu}_{(\chi)} \neq 0 $, the gravitational energy momentum tensor, $T^{\mu \nu}_{G}$ can vanish.
In this case the expanding matter produces no curvature of spacetime, and no curvature singularities are associated with the big bang of the matter. The geometry that contains the correct set of comoving observers is the Milne Universe and the matter is a gas of strings or K-matter.
As a model for the early universe, this solution is also free of the horizon problem, because cosmologies with a linear dependence in time for the Robertson Walker factor, as we have here, provide an infinite horizon distance (as in the Kolb coasting cosmology \cite{Kolb}).

The best definition of the energy of an object in an asymptotically flat spacetime is according to the gravitational effects it produces and the energy then equals the mass parameter in the asymptotic expansion of  $g_{00}$. According to this, the energy cost of producing these Minkowski flat universes is zero, although they may be filled with matter.  One can contemplate for example the possibility of constructing bubbles of material producing no gravitational energy momentum tensor and these bubbles could be therefore responsible for the creation of universes out of a pre-existing universe at no energy cost \cite{no energy cost}. There is also the interesting effect of "gravitational suppression" for some
kind of matter that has been found here. Gravitational suppression has been discussed phenomenologically in refs \cite{gravitational suppression}

The extension of the class of available solutions of this theory is of course very important and this has been done in the restricted form of the theory which is the TMT case but needs to be generalized.

Although we have not studied the theory in its most general form, coupling  $\chi_\mu$ to matter fields and even to curvature, study the theory in the framework of the first order formalism, as we have done with the special case of the theory which is the TMT, introduce supersymmetry (this seems possible), etc.,  but only started our investigations in this direction. Nevertheless, we have seen already what the basic features of any theory based on the introduction of a vector field that enforces the conservation law of a given energy momentum tensor, finds itself with another energy momentum tensor appearing, the gravitational energy momentum tensor and that the vector field  $\chi_\mu$
has the basic features of a "dynamical spacetime". The theory could also preserve the advantages found in its special form, the TMT case, in connection with the cosmological constant problem.

Finally, the availability of a dynamical time, instead of just a time parameter, could be of use in the context of quantum cosmology, where the notorious "time problem" has been discussed \cite{time problem}. This is exactly related to the fact that time in generally coordinate invariant theories is a parameter, totally arbitrary, and the Hamiltonian is therefore zero as a conequence of the time reparametrization invariance. This appears to produce two effects: i) time independence of physical states, ii) non normalizable character of physical states. Regularizing the physical states may force us out of the space of states which are time independent also and a resolution of the problem of time may be available this way \cite{time problem 2}, such a procedure can be carried out only in special cases so far.

There has been also an interesting proposal by Page and Wooters  \cite{Page} which is to promote some dynamical variable as the "clock"
and use this instead of the coordinate time.
It can be in general a dangerous approach since a wrong variable used in this way could lead to a great deal of confusion instead of solving something. 
Here is where the alternative use of a dynamical time, as we have defined here, instead of the coordinte time could be of use. $\chi^{\mu}$ is a vector whose zero  component has the features of a dynamical time, being even proportional to the Minkowski time coordinate when considering flat space time solutions. 
The consideration of $\chi^0$ in quantum gravity could therefore provide meaning to the concept of  time evolution in the context of the full quantum theory. Since although the time coordinate will still not enter into the wave function, instead $\chi^0$ will be entering in the solution of the functional equation of quantum gravity as one of the fields on which the wavefunction of the Universe depends, but being also the field which represents the dynamical time. In future publications we will study this proposal and the many technical questions associated to this
that will be necessary to to deal with in order to make out of this idea a viable solution to the time problem in quantum cosmology.

\section{Acknowledgements}
I would like to thank the group of Astrophysics and Cosmology of the Pontificia Universidad Catolica de Valparaiso, Chile, where this work was carried out during a sabbatical leave from Ben Gurion University, for hospitality and many physics discussions, in particular to Sergio del Campo, Ramon Herrera and Joel Saavedra. I also want to thank Luis A. Cabral for useful discussions concerning Killing Tensors, to Rodrigo Olea for useful conversations and discussions about symmetries in curved spacetime backgrounds, to Samuel Lepe for discussions concerning the physical implications of the solutions described here, to  Marcelo Samuel Berman and Luis Augusto Trevisan for conversations concerning the special features of the 
$ p =-\frac{1}{3}\rho $ equation of state. Finally I want also to thank Jacob Bekenstein for many interesting comments on the manuscript as well as many interesting physical insights.

\end{document}